\documentclass[aps,pra,superscriptaddress,twocolumn]{revtex4}
\usepackage[utf8x]{inputenc}
\usepackage{float}
\usepackage{bm}
\usepackage{graphicx,amsmath,latexsym,amssymb}
\usepackage{color}
\usepackage{dcolumn}
\begin{document}
\title{
Two-atom van-der-Waals forces with one atom excited: the identical atoms limit I} 
\author{J. S\'anchez-C\'anovas} 
\affiliation{Departamento de F\'isica Te\'orica, At\'omica y \'Optica and IMUVA,  Universidad de Valladolid, Paseo Bel\'en 7, 47011 Valladolid, Spain}
\author{M. Donaire} 
\email{manuel.donaire@uva.es}
\affiliation{Departamento de F\'isica Te\'orica, At\'omica y \'Optica and IMUVA,  Universidad de Valladolid, Paseo Bel\'en 7, 47011 Valladolid, Spain}

\begin{abstract}
We compute the conservative van-der-Waals forces between two atoms, one of which is initially excited, in the limit of  identical atoms.  
Starting with the perturbative calculation of the interaction between two dissimilar atoms, we show that a time-dependent approach in the weak-interaction approximation  
is essential in considering the identical atoms limit in the perturbative regime. In this limit we find that, 
at leading order, the van-der-Waals forces are fully-resonant and grow linearly in time, being different upon each atom.  
 The resultant net force upon the two-atom system 
 is related to the directionality of spontaneous emission, which results from the violation of parity symmetry. 
 In contrast to the usual stationary van-der-Waals forces, the time-dependent conservative forces cannot be written as the gradients of the expectation values 
 of the interaction potentials, but as the expectation values of the gradients of the interaction potentials only.
   

\end{abstract}
\maketitle
\section{Introduction} 
Dispersion forces between neutral atoms are the result of the coupling of the quantum fluctuations of the electromagnetic (EM) field in its vacuum state with the 
 fluctuations of the atomic charges in stable or metastable states. Generically, the corresponding forces are known as van-der-Waals (vdW) forces 
 \cite{Milonnibook,Craigbook,Miltonbook,BuhmannbookI,WileySipe}. In the last decades a 
 renewed interest has been drawn on the interaction between excited atoms. The interests are twofold. From a practical perspective, this is the kind of interaction 
 between Rydberg atoms \cite{Haroche1,Reinhard2007,Forster,Weber,Scheel,somepapers,Beguin} which makes possible the coherent manipulation of their quantum states, 
 facilitating the entanglement between 
 separated quantum systems as well as the storage of quantum information \cite{Cirac1,Cirac2,Zhen,Haroche2,ReviewHarocheetal,Saffman}. On the other hand, from a 
 fundamental perspective, the attention has 
 focused on different aspects of the interaction, namely, its scaling behavior with the distance \cite{SafariPRL,Berman,MePRL,Milonni,MePRA,Pablo,Jentschura1}, 
 the role of dissipation \cite{McLone,Power1965,Power1995,MePRA,Pablo}, its inherent time-dependence 
 \cite{Sherkunov1,Passante,Berman,MePRL,MePRA,Jentschura2}, and the net forces induced by parity and time-reversal violation \cite{My_Net_PRA,Sherkunov2}. 
 
 Hereafter and for the sake of simplicity we will consider the interaction between a pair of 
 two-level atoms, $A$ and $B$, with resonance frequencies $\omega_{A}$ and $\omega_{B}$, natural linewidths $\Gamma_{A}$ and $\Gamma_{B}$, and ground and 
 excited states labeled with subscripts $+$ and $-$, respectively, $|A_{\pm},B_{\pm}\rangle$. In the case of 
 dissimilar atoms, i.e., for $|\Delta_{AB}|=|\omega_{A}-\omega_{B}|\gg\Gamma_{A},\Gamma_{B}$, it is possible to use quasi-stationary perturbation theory to compute 
 the interaction. This is so because the excitation process can be taken adiabatic with respect to the rate at which the excitation is transferred between the 
 atoms, $\Delta_{AB}$. That is, denoting by $\Omega$ the Rabi frequency of the external exciting field, an adiabatic excitation holds for $|\Delta_{AB}|\gg\Omega$.  
 It was shown in Ref.\cite{MePRA} that, for arbitrary values of $\Omega$,  the resultant resonant interaction contains a quasi-stationary term which 
 oscillates in space with wavelength 
 $c\pi/\omega_{A}$ and is exponentially attenuated in time at the rate $\Gamma_{A}$, and time-oscillating terms of frequency $\Delta_{AB}$ whose amplitude is proportional to 
 $\Omega^{2}/(\Delta_{AB}^{2}-\Omega^{2})$. In the adiabatic limit the latter term vanishes \cite{Berman}, and the result is equivalent to that obtained 
 using adiabatic time-dependent perturbation theory \cite{MePRA}.  Other approaches based on Heisenberg's formalism \cite{Passante,Milonni,Pablo} 
and Feynman's Lagrangian formalism between asymptotic states \cite{Sherkunov1,Jentschura2} lead to an equivalent quasi-stationary result. In the opposite limit, 
that is, for a sudden  excitation with $\Omega\gg|\Delta_{AB}|$, quasi-stationary and time oscillating terms happen to be of the same order \cite{MePRL}. Either way, it was also found in Refs.\cite{MePRA,My_Net_PRA} that 
 a weak net force acts upon the center of mass of the two-atom system while  excited.
 
 As for the interaction of a binary system of identical two-level atoms, with one of them initially excited, neither quasi-stationary nor adiabatic approximations make physical sense
 for two reasons. In the first place, the system becomes degenerate, as the states $|A_{+},B_{-}\rangle$ and $|A_{-},B_{+}\rangle$ posses identical energies, and the 
 stationary sates are the symmetric and antisymmetric Dicke's states, $(|A_{+},B_{-}\rangle\pm|A_{-},B_{+}\rangle)/\sqrt{2}$, respectively. This implies 
 that the use of stationary perturbation theory becomes unsuitable. 
Second, in contrast to the interaction between dissimilar atoms, the null value of $\Delta_{AB}$ makes an adiabatic excitation unfeasible with respect to the 
original detuning. On the contrary, a sudden excitation is suitable as long as its associated Rabi frequency $\Omega$ is much greater than the detuning between 
the stationary Dicke's states \cite{Dicke,Craigbook}.  


In this article we will show that, starting with a binary system of dissimilar atoms, the identical atoms limit upon the interaction of the excited system 
can be formulated in a consistent manner using  
time-dependent perturbation theory in the sudden excitation approximation. In order to keep the calculation perturbative, we 
will restrict ourselves to the weak-interaction regime, meaning that the observation time is small in comparison to the time it takes for the excitation to be 
transferred from the initially 
excited atom to the other, which is of the order of the inverse of the detuning between the Dicke's states.  
We will show that, for two-level atoms, the van-der-Waals forces are dominated by fully-resonant components which 
  grow linearly in time and are different upon each atom. Besides, in addition to the familiar off-resonant van-der-Waals force, 
  a reciprocal semi-resonant force arises. 
Interestingly,  the time-dependent forces do not derive from the gradients of the expectation values of the 
 interaction potentials, but from the expectation values of the gradients of the interaction potentials instead. 
  The non-reciprocal components of the force are explained in terms of parity symmetry violation, which  generate an asymmetry in the probability of emission of photons from either atom.
    The effect of the de-excitation upon the off-resonant van-der-Waals force is also analyzed. 
    
The article is organized as follows. In Sec.\ref{sec2} we perform the computation of the vdW forces between two 
dissimilar two-level atoms, one of which is suddenly excited. The origin of non-reciprocal forces 
is related to the directionality of spontaneous emission. In Sec.\ref{sec3} the identical atoms limit is considered in the weak-interaction regime. 
The conclusions are summarized in Sec.\ref{sec4} together with a discussion on the extension of our results. 

\section{vdW interaction of two dissimilar atoms after a sudden excitation}\label{sec2}

Let us consider two atoms, $A$ and $B$, located a distance $R$ apart. Since we are ultimately interested in the identical atoms limit, 
$|\Delta_{AB}|\ll\Gamma_{A}$, $\Gamma_{A}\rightarrow\Gamma_{B}$, atom $A$ is assumed to be suddenly excited with an external 
field of strength $\Omega\gg|\Delta_{AB}|$. This is the situation considered in Ref.\cite{MePRL}, where the calculation was restricted to  
quasi-resonant processes, and to observation times $T$ such that $\Gamma_{A,B}T\ll1$. Here we will go beyond those restrictions and we will evaluate all the 
contributions to the vdW forces on both atoms, at leading order in the coupling parameter.

Let us consider a sudden excitation of atom $A$.  The state of the system at time 0 is $|\Psi(0)\rangle=|A_{+}\rangle\otimes|B_{-}\rangle\otimes|0_{\gamma}\rangle$, 
where  $(A,B)_{\pm}$ label the upper/lower internal states of the atoms $A$ and $B$ respectively, and $|0_{\gamma}\rangle$ is the electromagnetic (EM) vacuum state. At any given time $T>0$ the state of the two-atom-EM field system can be written as $|\Psi(T)\rangle=\mathbb{U}(T)|\Psi(0)\rangle$, where $\mathbb{U}(T)$ denotes the time propagator in the Schr\"odinger representation,
\begin{align}
\mathbb{U}(T)&=\mathcal{T}\exp{}\Bigl\{-i\hbar^{-1}\int_{0}^{T}\textrm{d}t\:H\Bigr\},\label{UT}\\
H&=H_{A}+H_{B}+H_{EM}+W.\nonumber
\end{align}
In this equation $H_{A}+H_{B}$ is the free Hamiltonian of the internal atomic states, 
$\hbar\omega_{A}|A_{+}\rangle\langle A_{+}|+\hbar\omega_{B}|B_{+}\rangle\langle B_{+}|$, while the Hamiltonian of the free EM field is 
$H_{EM}=\sum_{\mathbf{k},\boldsymbol{\epsilon}}\hbar\omega(a^{\dagger}_{\mathbf{k},\boldsymbol{\epsilon}}a_{\mathbf{k},\boldsymbol{\epsilon}}+1/2)$,
where $\omega=ck$ is the photon frequency, and the operators $a^{\dagger}_{\mathbf{k},\boldsymbol{\epsilon}}$ and $a_{\mathbf{k},\boldsymbol{\epsilon}}$ are the creation 
and annihilation operators of photons with momentum $\hbar\mathbf{k}$ and polarization $\boldsymbol{\epsilon}$, respectively. Finally, the interaction Hamiltonian in 
the electric dipole approximation reads $W=W_{A}+W_{B}$, with 
\begin{eqnarray}
W_{A,B}&\simeq&-\mathbf{d}_{A,B}\cdot\mathbf{E}(\mathbf{R}_{A,B}).\label{WAB}
\end{eqnarray}
 In this expression $\mathbf{d}_{A,B}$ are the electric dipole operators of each atom, and $\mathbf{E}(\mathbf{R}_{A,B})$ is the quantum electric field operators 
 in Schr\"odinger's representation evaluated at 
the position of the center of mass of each atom, $\mathbf{R}_{A,B}$, respectively. In terms of the EM vector potential,
\begin{equation}
\mathbf{A}(\mathbf{r},t)=\sum_{\mathbf{k},\boldsymbol{\epsilon}}\sqrt{\frac{\hbar}{2\omega\mathcal{V}\epsilon_{0}}}
[\boldsymbol{\epsilon}a_{\mathbf{k},\boldsymbol{\epsilon}}e^{i(\mathbf{k}\cdot\mathbf{r}-\omega t)}
+\boldsymbol{\epsilon}^{*}a^{\dagger}_{\mathbf{k},\boldsymbol{\epsilon}}e^{-i(\mathbf{k}\cdot\mathbf{r}-\omega t)}],\nonumber
\end{equation}
 the electric field $\mathbf{E}(\mathbf{R}_{A,B})=-\partial_{t}\mathbf{A}(\mathbf{R}_{A,B},t)|_{t=0}$ can be written as a sum over normal modes,
\begin{eqnarray}\label{AQ}
\mathbf{E}(\mathbf{R}_{A,B})&=&\sum_{\mathbf{k}}\mathbf{E}^{(-)}_{\mathbf{k}}(\mathbf{R}_{A,B})+\mathbf{E}^{(+)}_{\mathbf{k}}(\mathbf{R}_{A,B})\nonumber\\
&=&i\sum_{\mathbf{k},\boldsymbol{\epsilon}}\sqrt{\frac{\hbar ck}{2\mathcal{V}\epsilon_{0}}}
[\boldsymbol{\epsilon}a_{\mathbf{k},\boldsymbol{\epsilon}}e^{i\mathbf{k}\cdot\mathbf{R}_{A,B}}-\boldsymbol{\epsilon}^{*}a^{\dagger}_{\mathbf{k},\boldsymbol{\epsilon}}e^{-i\mathbf{k}\cdot\mathbf{R}_{A,B}}],\nonumber
\end{eqnarray}
where $\mathcal{V}$ is a generic volume and $\mathbf{E}^{(\mp)}_{\mathbf{k}}$ denote the annihilation/creation electric field operators of photons of momentum 
$\hbar\mathbf{k}$, respectively. Strictly speaking, $W$ includes an additional term in the electric dipole approximation which is referred to 
as  R\"ontgen term \cite{Baxter1_Baxter2}. As argued in Ref.\cite{MeQFriction}, that term is negligible since its contribution 
to Eq.(\ref{UT}) contains terms of orders $\dot{R}_{A,B}/c$ and $\mathbf{d}_{A,B}\cdot\mathbf{E}(\mathbf{R}_{A,B})/m_{A,B}$ smaller than the contribution of 
Eq.(\ref{WAB}), with $m_{A,B}$ being the atomic masses.

Next, considering $W$ as a perturbation to the free Hamiltonians, the unperturbed time propagator for atom and free photon states is $\mathbb{U}_{0}(t)=\exp{[-i\hbar^{-1}(H_{A}+H_{B}+H_{EM})t]}$. In terms of $W$ and $\mathbb{U}_{0}$, $\mathbb{U}(T)$ admits an expansion in powers of  $W$ which can be developed out of the time-ordered exponential equation,
\begin{equation}\label{U}
\mathbb{U}(T)=\mathbb{U}_{0}(T)\:\mathcal{T}\exp\int_{0}^{T}(-i/\hbar)\mathbb{U}_{0}^{\dagger}(t)\:W\:\mathbb{U}_{0}(t)\textrm{d}t,
\end{equation}
which can be written as a series in powers of  $W$ as $\mathbb{U}(T)=\mathbb{U}_{0}(T)+\sum_{n=1}^{\infty}\delta\mathbb{U}^{(n)}(T)$, with $\delta\mathbb{U}^{(n)}$ being the term of order $W^{n}$. 

The system posseses a conserved total momentum  \cite{Cohen_Kawkathesis_KawkavanTiggelen_Dippel,Kawkathesis}, $[H,\mathbf{K}]=\mathbf{0}$, 
\begin{equation}
\mathbf{K}=\mathbf{P}_{A}+\mathbf{P}_{B}+\mathbf{P}_{\perp}^{\gamma},\label{K}
\end{equation}
where $\mathbf{P}_{A,B}$ are the  canonical conjugate momenta of the centers of mass of each atom and $\mathbf{P}_{\perp}^{\gamma}=\sum_{\mathbf{k},\mathbf{\epsilon}}\hbar\mathbf{k}\:a^{\dagger}_{\mathbf{k},\mathbf{\epsilon}}a_{\mathbf{k},\mathbf{\epsilon}}$ 
is the transverse EM momentum. Further, if the charges  $\{q_{i}\}$ within the atoms are considered individually at positions $\{\mathbf{r}_{i}\}$, the canonical conjugate 
momenta can be written as 
\begin{equation}
\mathbf{P}_{A}+\mathbf{P}_{B}=m_{A}\dot{\mathbf{R}}_{A}+m_{B}\dot{\mathbf{R}}_{B}+\sum_{i}q_{i}\mathbf{A}(\mathbf{r}_{i}),\label{conjugate}
\end{equation}
where the first two terms are the kinetic momenta of the centers of mass of each atom, and the momentum within the summation symbol is referred to as longitudinal EM momentum \cite{Cohen_Kawkathesis_KawkavanTiggelen_Dippel}, 
$\mathbf{P}_{\parallel}^{\gamma}=\sum_{i}q_{i}\mathbf{A}(\mathbf{r}_{i})$. Lastly, in the electric dipole approximation, $\mathbf{P}_{\parallel}^{\gamma}$ reads 
\cite{Baxter1_Baxter2}, $\mathbf{P}_{\parallel}^{\gamma}\simeq-\mathbf{d}_{A}\times\mathbf{B}(\mathbf{R}_{A})-\mathbf{d}_{B}\times\mathbf{B}(\mathbf{R}_{B})$, 
where $\mathbf{B}(\mathbf{R}_{A,B})=\boldsymbol{\nabla}_{A,B}\times\mathbf{A}(\mathbf{R}_{A,B})$.

Following Refs.\cite{MeQFriction,My_Net_PRA}, the force on each atom is computed applying the time derivative to the expectation value of the kinetic momenta 
of the centers of mass of each atom. Writing the latter in terms of the canonical conjugate momenta and the longitudinal EM momentum, we arrive at
\begin{align}
\langle\mathbf{F}_{A,B}\rangle_{T}&=\partial_{T}\langle m_{A,B}\dot{\mathbf{R}}_{A,B}\rangle_{T}\label{Force}\\
&=-i\hbar\partial_{T}\langle\Psi(0)|\mathbb{U}^{\dagger}(T)\boldsymbol{\nabla}_{A,B}\mathbb{U}(T)|\Psi(0)\rangle\nonumber\\
&+\partial_{T}\langle\Psi(0)|\mathbb{U}^{\dagger}(T)\mathbf{d}_{A,B}\times\mathbf{B}(\mathbf{R}_{A,B})\mathbb{U}(T)|\Psi(0)\rangle\nonumber\\
&=-\langle \boldsymbol{\nabla}_{A,B}W_{A,B}\rangle_{T}+
\partial_{T}\langle\mathbf{d}_{A,B}\times\mathbf{B}(\mathbf{R}_{A,B})\rangle_{T},\nonumber
\end{align}
The first term on the right hand side of last equality is a conservative force along the interatomic axis, which we will refer to as vdW force. Note hower that, in contrast to the stationary vdW forces 
 computed in the adiabatic approximation --cf. Ref.\cite{MePRA}, time-dependent conservative forces cannot be generally written as 
$\frac{-1}{2}\boldsymbol{\nabla}_{A,B}\langle W_{A,B}\rangle_{T}$. We will show latter, including up to two-photon exchange processes, that 
the reason is the functional dissymmetry in the contribution of the two photons to the time-dependent terms.
The second term is a non-conservative force equivalent to the time derivative of the longitudinal EM momentum at each atom, with opposite sign. We will 
show in a separate publication \cite{PRAJulio2} that its contribution is only observable for $|\Delta_{AB}|\ll\omega_{A,B}$, being of the order of 
max$(|\Delta_{AB}|,\Gamma_{A})/\omega_{A}$ times smaller than the vdW conservative force. Hereafter we will neglect it and approximate  
$\langle\mathbf{F}_{A,B}\rangle_{T}\simeq-\langle \boldsymbol{\nabla}_{A,B}W_{A,B}\rangle_{T}$.

\begin{figure}[H]
\includegraphics[width=8.5cm,clip]{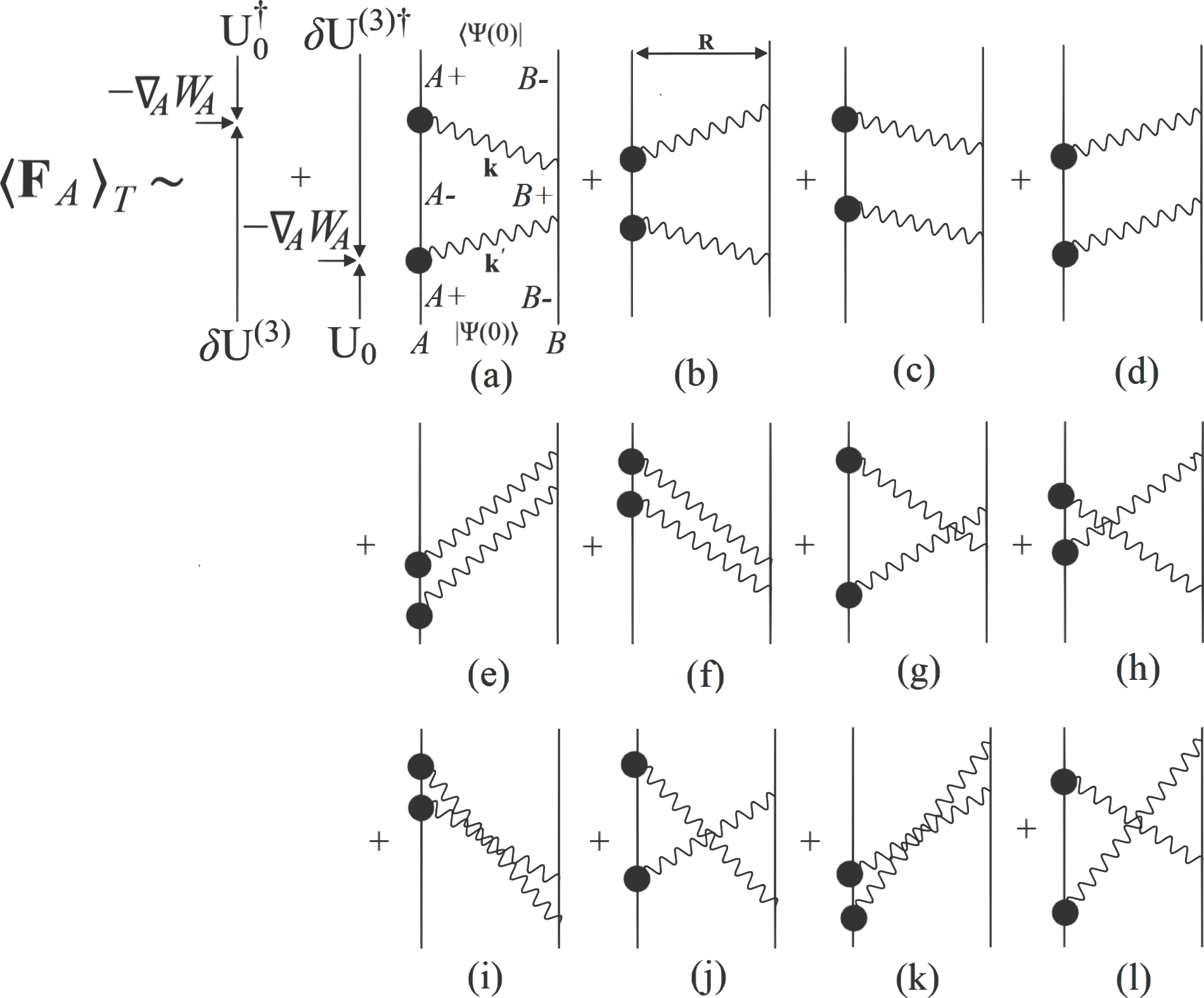}
\caption{Diagrammatic representation of twelve of the processes which contribute to $\langle\mathbf{F}_{A}\rangle_{T}$. 
Solid straight lines stand for propagators of atomic states, while wavy lines stand for photon propagators. In diagram (a), atomic and photon states are indicated  
explicitly. The atoms $A$ and $B$ are separated by a distance $R$ along the horizontal direction, whereas time runs along the vertical. 
The big circles in black on the left of each diagram stand for the insertion of the Schr\"odinger operator $-\boldsymbol{\nabla}_{A}W_{A}$ whose expectation value is computed.  
Each diagram contributes with two terms, one from each of the operators inserted. They are sandwiched between two time propagators, $\textrm{U}(T)$ and 
$\textrm{U}^{\dagger}(T)$ (depicted by vertical arrows), which evolve the initial state $|\Psi(0)\rangle$ towards the observation time at which $-\boldsymbol{\nabla}_{A}W_{A}$ 
applies.}\label{figure1A}
\end{figure}

\begin{figure}[H]
\includegraphics[
width=8.5cm,clip]{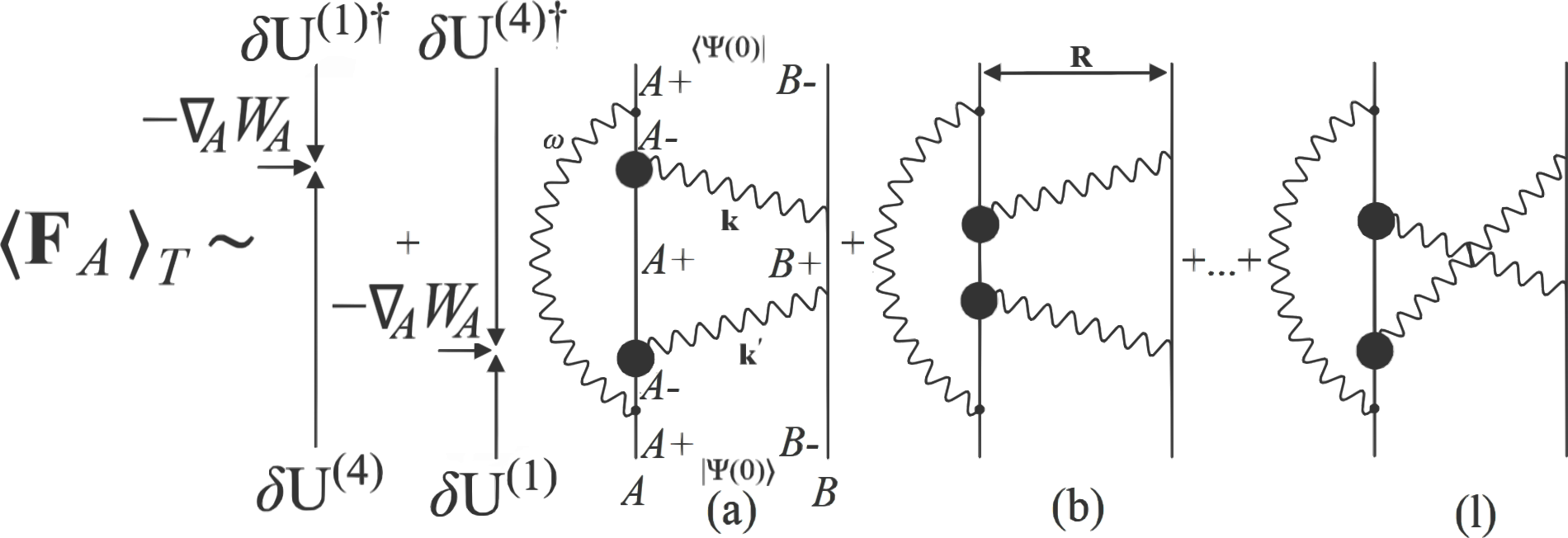}
\caption{Diagrammatic representation of processes which contribute to the fully off-resonant component of $\langle\mathbf{F}_{A}\rangle_{T}$. 
In contrast to the diagrams in Fig.\ref{figure1A}, the self-interacting photon on atom $A$ leads the spontaneous emission from the excited atom. 
The omitted diagrams are analogous to those in Fig.\ref{figure1A}.}\label{figure2A}
\end{figure}

A perturbative development of Eq.(\ref{Force}) shows that, up to terms involving two-photon exchange processes,  twenty-four diagrams contribute to 
$\langle \mathbf{F}_{A}\rangle_{T}$ for a two-level atom. They are depicted in 
Fig.\ref{figure1A} and Fig.\ref{figure2A}. Note that those in Fig.\ref{figure2A} just differ with respect to those of Fig.\ref{figure1A} by the photon embracing 
the two exchanged photons, which accounts for the de-excitation of the system via spontaneous emission from atom $A$.
For the sake of illustration we give below the expression of diagram (a) in Fig.\ref{figure1A}, which contributes to $\langle\mathbf{F}_{A}\rangle_{T}$ in the form,
\begin{widetext}
\begin{align}\label{laeq}
&\frac{1}{\hbar^{3}}\int_{0}^{\infty}\frac{\mathcal{V}k^{2}\textrm{d}k}{(2\pi)^{3}}\int_{0}^{\infty}\frac{\mathcal{V}k^{'2}\textrm{d}k'}{(2\pi)^{3}}\int_{0}^{4\pi}\textrm{d}\Theta\int_{0}^{4\pi}\textrm{d}\Theta'\Bigl\{\Bigl[i\langle A_{+},B_{-},0_{\gamma}|e^{i \Omega_{a}^{*}T}| A_{+},B_{-},0_{\gamma}\rangle\int_{-\infty}^{T}\textrm{d}t\int_{-\infty}^{t}\textrm{d}t'\int_{-\infty}^{t'}\textrm{d}t''\nonumber\\
&\times\:\langle  A_{+},B_{-},0_{\gamma}|-\boldsymbol{\nabla}_{A}[\mathbf{d}_{A}\cdot\mathbf{E}_{\mathbf{k}'}^{(-)}(\mathbf{R}_{A})]|A_{-},B_{-},\gamma_{k'}\rangle e^{-i\omega'(T-t)}\langle A_{-},B_{-},\gamma_{k'}|\mathbf{d}_{B}\cdot\mathbf{E}_{\mathbf{k}'}^{(+)}(\mathbf{R}_{B})|A_{-},B_{+},0_{\gamma}\rangle\nonumber\\
&\times e^{-i \Omega_{b}(t-t')}\langle A_{-},B_{+},0_{\gamma}|\mathbf{d}_{B}\cdot\mathbf{E}_{\mathbf{k}}^{(-)}(\mathbf{R}_{B})|A_{-},B_{-},\gamma_{k}\rangle e^{-i\omega(t'-t'')}\langle A_{-},B_{-},\gamma_{k}|\mathbf{d}_{A}\cdot\mathbf{E}_{\mathbf{k}}^{(+)}(\mathbf{R}_{A})| A_{+},B_{-},0_{\gamma}\rangle e^{-i \Omega_{a} t''}\Bigr]\nonumber\\
&+[k\leftrightarrow k']^{\dagger}\Bigr\},
\end{align} 
\end{widetext}
where it is implicit that the causality condition $T\gg R/c$ holds at the time of observation. In this equation $|A_{+},B_{-},0_{\gamma}\rangle$ is the initial 
two-atom-EM-vacuum state, with atom $A$ excited at time 0, $|\gamma_{k}\rangle$ is a one-photon state of momentum $\mathbf{k}$ and frequency $\omega=ck$, the 
complex time-exponentials are the result of the application of the free time-evolution operator $\textrm{U}_{0}(t)=e^{-i\hbar^{-1}H_{0}t}$ between the interaction 
vertices $W_{A,B}$, with $ \Omega_{a}=\omega_{A}-i\Gamma_{A}/2$ and $ \Omega_{b}=\omega_{B}-i\Gamma_{B}/2$, where the 
dissipative imaginary terms account for radiative emission in the Weisskopf-Wigner approximation. After integrating in time and solid angles,  one arrives at 
\begin{widetext}
\begin{align}
&\frac{c^{2} \hbar^{-1}}{\pi^{2}\epsilon_{0}^{2}}\textrm{Re}\int_{0}^{\infty}\textrm{d}k'k^{'2}\boldsymbol{\nabla}_{A}[\boldsymbol{\mu}_{A}\cdot\textrm{Im}\mathbb{G}(k'R)\cdot
\boldsymbol{\mu}_{B}]\int_{0}^{\infty}\textrm{d}k\:k^{2}\boldsymbol{\mu}_{B}\cdot\textrm{Im}\mathbb{G}(kR)\cdot\boldsymbol{\mu}_{A}\:e^{i \Omega_{a}^{*}T}\Bigl[ \frac{e^{-i  \Omega_{a} T}-e^{-i \omega T}}{(\omega'- \Omega_{a})( \Omega_{b}- \Omega_{a})(\omega -  \Omega_{a})}\nonumber\\
&-	\frac{e^{-i  \Omega_{b} T}-e^{-i \omega T}}{(\omega'- \Omega_{a})( \Omega_{b}- \Omega_{a})(\omega -  \Omega_{b})}+\frac{e^{-i \omega' T}-e^{-i \omega T}}{(\omega'- \Omega_{a})(\omega'- \Omega_{b})(\omega - \omega')}-\frac{e^{-i  \Omega_{b} T}-e^{-i \omega T}}{(\omega'- \Omega_{a})(\omega'- \Omega_{b})(\omega -  \Omega_{b})} \Bigr]
\end{align}
\end{widetext}

\noindent where $\boldsymbol{\mu}_{A}=\langle A_{-}|\mathbf{d}_{A}|A_{+}\rangle$, $\boldsymbol{\mu}_{B}=\langle B_{-}|\mathbf{d}_{B}|B_{+}\rangle$ and $\mathbb{G}(kR)$ is 
the dyadic Green's function of the electric field induced at $\mathbf{R}$ by an electric dipole of frequency $\omega=ck$ placed at the origin. It reads
\begin{equation}
\mathbb{G}(kR)=\frac{k\:e^{ikR}}{-4\pi}[ \alpha/kR+i \beta/(kR)^{2}- \beta/(kR)^{3}],
\end{equation}
where the tensors $ \alpha$ and $ \beta$ read $ \alpha=\mathbb{I}-\mathbf{R}\mathbf{R}/R^{2}$,  $ \beta=\mathbb{I}-3\mathbf{R}\mathbf{R}/R^{2}$. 

Operating in an analogous fashion with the rest of the terms derived from the diagrams of Figs.\ref{figure1A} and \ref{figure2A}, upon integration 
in $k$ and $k'$ in the complex plane, using the identity $\boldsymbol{\nabla}_{B}=-\boldsymbol{\nabla}_{A}=-\boldsymbol{\nabla}_{\mathbf{R}}$, we arrive at
\begin{widetext}
\begin{equation}
\begin{split}
&\langle \mathbf{F}_A\rangle_{T}=-\frac{2\omega_A^4 e^{-\Gamma_A T}}{c^4 \epsilon_0^2\hbar\Delta_{AB}}\Bigl[\boldsymbol{\mu}_{A}\cdot\textrm{Re}\mathbb{G}(k_A R)\cdot\boldsymbol{\mu}_{B}
\boldsymbol{\nabla}_{\mathbf{R}}\left[\boldsymbol{\mu}_{B}\cdot\textrm{Re}\mathbb{G}(k_A R)\cdot\boldsymbol{\mu}_{A}\right]-\boldsymbol{\mu}_{A}\cdot\textrm{Im}
\mathbb{G}(k_A R)\cdot\boldsymbol{\mu}_{B}\boldsymbol{\nabla}_{\mathbf{R}}\left[\boldsymbol{\mu}_{B}\cdot\textrm{Im}\mathbb{G}(k_A R)\cdot\boldsymbol{\mu}_{A}\right]\Bigr]\\
&+\frac{2\omega_B^4 e^{-(\Gamma_A+\Gamma_{B})T/2}}{c^4 \epsilon_0^2\hbar\Delta_{AB}}\Bigl[\boldsymbol{\mu}_{A}\cdot\textrm{Re}\mathbb{G}(k_B R)\cdot\boldsymbol{\mu}_{B}
\boldsymbol{\nabla}_{\mathbf{R}}\left[\boldsymbol{\mu}_{B}\cdot\textrm{Re}\mathbb{G}(k_B R)\cdot\boldsymbol{\mu}_{A}\right]-\boldsymbol{\mu}_{A}\cdot\textrm{Im}
\mathbb{G}(k_B R)\cdot\boldsymbol{\mu}_{B}\boldsymbol{\nabla}_{\mathbf{R}}\left[\boldsymbol{\mu}_{B}\cdot\textrm{Im}\mathbb{G}(k_B R)\cdot\boldsymbol{\mu}_{A}\right]\Bigr]\\
&\times\cos(\Delta_{AB}T)\\
&-\frac{2\omega_B^4 e^{-(\Gamma_A+\Gamma_{B})T/2}}{c^4 \epsilon_0^2\hbar\Delta_{AB}}\Bigl[\boldsymbol{\mu}_{A}\cdot\textrm{Re}\mathbb{G}(k_B R)\cdot\boldsymbol{\mu}_{B}
\boldsymbol{\nabla}_{\mathbf{R}}\left[\boldsymbol{\mu}_{B}\cdot\textrm{Im}\mathbb{G}(k_B R)\cdot\boldsymbol{\mu}_{A}\right]+\boldsymbol{\mu}_{A}\cdot\textrm{Im}
\mathbb{G}(k_B R)\cdot\boldsymbol{\mu}_{B}\boldsymbol{\nabla}_{\mathbf{R}}\left[\boldsymbol{\mu}_{B}\cdot\textrm{Re}\mathbb{G}(k_B R)\cdot\boldsymbol{\mu}_{A}\right]\Bigr]\\
&\times\sin(\Delta_{AB}T)\\
&+\frac{2  \omega_A^4 e^{-\Gamma_A T}}{c^4 \epsilon_0^2 \hbar(\omega_A+\omega_B)}\Bigl[\boldsymbol{\mu}_{A}\cdot\textrm{Re}\mathbb{G}(k_A R)\cdot\boldsymbol{\mu}_{B}
\boldsymbol{\nabla}_{\mathbf{R}}\left[\boldsymbol{\mu}_{B}\cdot\textrm{Re}\mathbb{G}(k_A R)\cdot\boldsymbol{\mu}_{A}\right]-\boldsymbol{\mu}_{A}\cdot\textrm{Im}
\mathbb{G}(k_A R)\cdot\boldsymbol{\mu}_{B}\boldsymbol{\nabla}_{\mathbf{R}}\left[\boldsymbol{\mu}_{B}\cdot\textrm{Im}\mathbb{G}(k_A R)\cdot\boldsymbol{\mu}_{A}\right]\Bigr]\\
&-\frac{2 \omega_B^2  e^{-(\Gamma_A+\Gamma_B) T/2} }{c^3 \epsilon_0^2 \hbar}\Bigl[\boldsymbol{\nabla}_{\mathbf{R}}[\boldsymbol{\mu}_{A}\cdot\textrm{Re}\mathbb{G}(k_B R)
\cdot\boldsymbol{\mu}_B]\cos(\Delta_{AB}T)-\boldsymbol{\nabla}_{\mathbf{R}}[\boldsymbol{\mu}_{A}\cdot\textrm{Im}\mathbb{G}(k_B R)
\cdot\boldsymbol{\mu}_B]\sin(\Delta_{AB}T)\Bigr]\\
&\times \int_{0}^{\infty}\frac{dq}{\pi} \frac{(q^2-k_A k_B)q^2\boldsymbol{\mu}_{A}\cdot\mathbb{G}(iq R)\cdot\boldsymbol{\mu}_{B}}{(q^2+k_A^2)(q^2+k_B^2)}
+\frac{4\omega_A \omega_B\left(1-2e^{-\Gamma_{A} T}\right)}{c^3\epsilon_{0}^{2}\hbar}\int_{0}^{\infty}\frac{dq}{\pi}
\frac{q^4 \boldsymbol{\mu}_{A}\cdot\mathbb{G}(iqR)\cdot\boldsymbol{\mu}_{B} }{(q^2+k_A^2)(q^2+k_B^2)}
\boldsymbol{\nabla}_{\mathbf{R}}[\boldsymbol{\mu}_{B}\cdot\mathbb{G}(iqR)\cdot\boldsymbol{\mu}_{A}].\label{FAdist}
\end{split}
\end{equation}
\end{widetext}
In this equation, negligible and unobservable terms have been discarded. These are, off-resonant terms whose integrands are attenuated in time as $e^{-cTq}$ and whose contribution is $(R/cT)^{3}\ll1$ times smaller; and fast oscillating terms of frequency $\omega_{A}+\omega_{B}$ which average to zero upon observation. The 
origin of the terms in Eq.(\ref{FAdist}) is as follows. The first three terms, which scale as $\sim1/\Delta_{AB}$, are fully resonant and involve the evaluation of the 
two residues associated to simple poles in $k$ and $k'$ in the integrals stemming from diagram (a). The fourth term, which scales as $1/(\omega_{A}+\omega_{B})$, 
fully resonant too, results from the two resonant photons of diagram (g), which contains a two-photon intermediate state. The semi-resonant terms, which entail evaluating the residue associated to a simple pole in $k$ or $k'$ only, oscillate in time at frequency $\Delta_{AB}$. They stem from diagrams (c,d,e,f). Finally, the last term is the result of the addition of the 
 off-resonant contributions coming from the twelve diagrams of Figs.\ref{figure1A} and \ref{figure2A} together. The discarded fast oscillating terms, resonant and semi-resonant, are associated to diagrams (i,k) and (i,j,k,l), respectively, which contain two-photon intermediate states. 

\begin{figure}[H]
\includegraphics[width=8.5cm,clip]{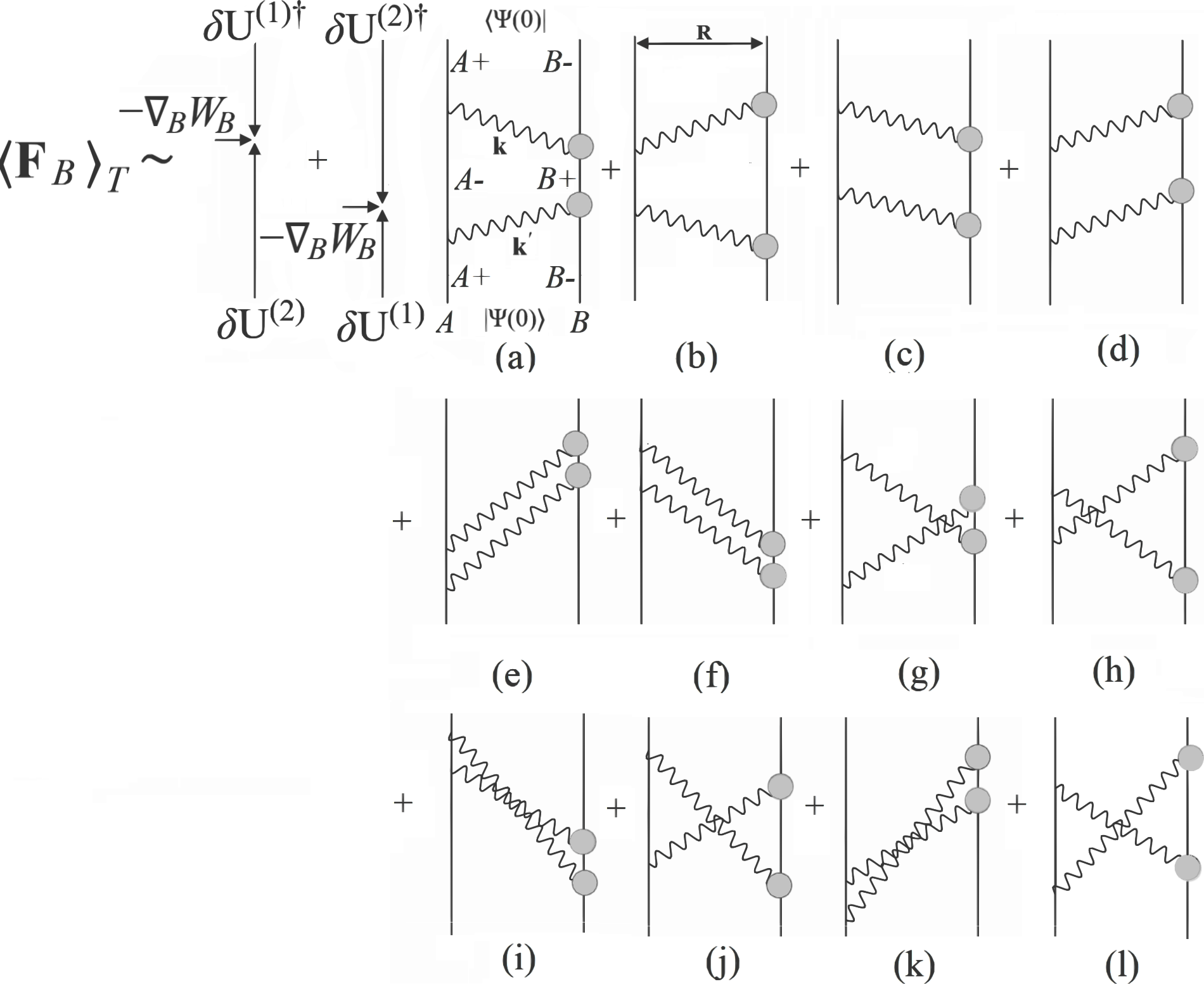}
\caption{Diagrammatic representation of twelve of the processes which contribute to $\langle\mathbf{F}_{B}\rangle_{T}$.  The big circles in gray on the right of 
each diagram stand for the insertion of the Schr\"odinger operator $-\boldsymbol{\nabla}_{B}W_{B}$ whose expectation value is computed.}\label{figure1B}
\end{figure}

\begin{figure}[H]
	\includegraphics[width=8.5cm,clip]{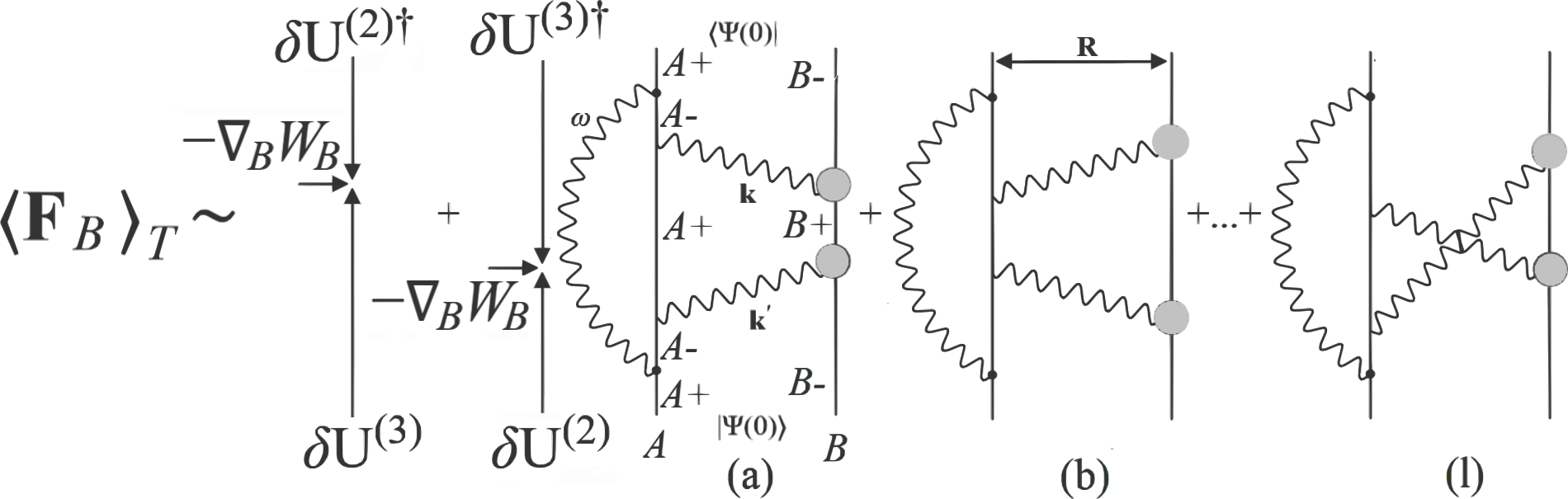}
	\caption{Diagrammatic representation of processes which contribute to the fully off-resonant component of $\langle\mathbf{F}_{B}\rangle_{T}$. 
		In contrast to the diagrams in Fig.\ref{figure1B}, the self-interacting photon on atom $A$ leads the spontaneous emission from the excited atom. 
		The omitted diagrams are analogous to those in Fig.\ref{figure1B}.}\label{figure2B}
\end{figure}

Analogous diagrams hold for $\langle\mathbf{F}_{B}\rangle_{T}$, but for the evaluation of the operator $-\boldsymbol{\nabla}_{B}W_{B}$ at atom $B$ 
--see Figs.\ref{figure1B} and \ref{figure2B},
\begin{widetext}
\begin{equation}
\begin{split}
&\langle \mathbf{F}_B\rangle_{T}=\frac{2\omega_A^4 e^{-\Gamma_A T}}{c^4 \epsilon_0^2\hbar\Delta_{AB}}\Bigl[\boldsymbol{\mu}_{A}\cdot\textrm{Re}\mathbb{G}(k_A R)\cdot\boldsymbol{\mu}_{B}
\boldsymbol{\nabla}_{\mathbf{R}}\left[\boldsymbol{\mu}_{B}\cdot\textrm{Re}\mathbb{G}(k_A R)\cdot\boldsymbol{\mu}_{A}\right]+\boldsymbol{\mu}_{A}\cdot\textrm{Im}
\mathbb{G}(k_A R)\cdot\boldsymbol{\mu}_{B}\boldsymbol{\nabla}_{\mathbf{R}}\left[\boldsymbol{\mu}_{B}\cdot\textrm{Im}\mathbb{G}(k_A R)\cdot\boldsymbol{\mu}_{A}\right]\Bigr]\\
&-\frac{2\omega_B^2\omega_{A}^{2} e^{-(\Gamma_A+\Gamma_{B})T/2}}{c^4 \epsilon_0^2\hbar\Delta_{AB}}\Bigl[\boldsymbol{\mu}_{A}\cdot\textrm{Re}\mathbb{G}(k_B R)\cdot\boldsymbol{\mu}_{B}
\boldsymbol{\nabla}_{\mathbf{R}}\left[\boldsymbol{\mu}_{B}\cdot\textrm{Re}\mathbb{G}(k_A R)\cdot\boldsymbol{\mu}_{A}\right]+\boldsymbol{\mu}_{A}\cdot\textrm{Im}
\mathbb{G}(k_B R)\cdot\boldsymbol{\mu}_{B}\boldsymbol{\nabla}_{\mathbf{R}}\left[\boldsymbol{\mu}_{B}\cdot\textrm{Im}\mathbb{G}(k_A R)\cdot\boldsymbol{\mu}_{A}\right]\Bigr]\\
&\times\cos(\Delta_{AB}T)\\
&-\frac{2\omega_B^2\omega_{A}^{2} e^{-(\Gamma_A+\Gamma_{B})T/2}}{c^4 \epsilon_0^2\hbar\Delta_{AB}}\Bigl[\boldsymbol{\mu}_{A}\cdot\textrm{Re}\mathbb{G}(k_B R)\cdot\boldsymbol{\mu}_{B}
\boldsymbol{\nabla}_{\mathbf{R}}\left[\boldsymbol{\mu}_{B}\cdot\textrm{Im}\mathbb{G}(k_A R)\cdot\boldsymbol{\mu}_{A}\right]-\boldsymbol{\mu}_{A}\cdot\textrm{Im}
\mathbb{G}(k_B R)\cdot\boldsymbol{\mu}_{B}\boldsymbol{\nabla}_{\mathbf{R}}\left[\boldsymbol{\mu}_{B}\cdot\textrm{Re}\mathbb{G}(k_A R)\cdot\boldsymbol{\mu}_{A}\right]\Bigr]\\
&\times\sin(\Delta_{AB}T)\\
&-\frac{2  \omega_A^4 e^{-\Gamma_A T}}{c^4 \epsilon_0^2 \hbar(\omega_A+\omega_B)}\Bigl[\boldsymbol{\mu}_{A}\cdot\textrm{Re}\mathbb{G}(k_A R)\cdot\boldsymbol{\mu}_{B}
\boldsymbol{\nabla}_{\mathbf{R}}\left[\boldsymbol{\mu}_{B}\cdot\textrm{Re}\mathbb{G}(k_A R)\cdot\boldsymbol{\mu}_{A}\right]+\boldsymbol{\mu}_{A}\cdot\textrm{Im}
\mathbb{G}(k_A R)\cdot\boldsymbol{\mu}_{B}\boldsymbol{\nabla}_{\mathbf{R}}\left[\boldsymbol{\mu}_{B}\cdot\textrm{Im}\mathbb{G}(k_A R)\cdot\boldsymbol{\mu}_{A}\right]\Bigr]\\
&+\frac{2 \omega_A^2  e^{-(\Gamma_A+\Gamma_B) T/2} }{c^3 \epsilon_0^2 \hbar}\Bigl[\boldsymbol{\nabla}_{\mathbf{R}}[\boldsymbol{\mu}_{A}\cdot\textrm{Re}\mathbb{G}(k_A R)
\cdot\boldsymbol{\mu}_B]\cos(\Delta_{AB}T)+\boldsymbol{\nabla}_{\mathbf{R}}[\boldsymbol{\mu}_{A}\cdot\textrm{Im}\mathbb{G}(k_A R)
\cdot\boldsymbol{\mu}_B]\sin(\Delta_{AB}T)\Bigr]\\
&\times \int_{0}^{\infty}\frac{dq}{\pi} \frac{(q^2-k_A k_B)q^2\boldsymbol{\mu}_{A}\cdot\mathbb{G}(iq R)\cdot\boldsymbol{\mu}_{B}}{(q^2+k_A^2)(q^2+k_B^2)}-\frac{4\omega_A \omega_B\left(1-2e^{-\Gamma_{A} T}\right)}{c^3\epsilon_{0}^{2}\hbar}\int_{0}^{\infty}\frac{dq}{\pi}
\frac{q^4 \boldsymbol{\mu}_{A}\cdot\mathbb{G}(iqR)\cdot\boldsymbol{\mu}_{B} }{(q^2+k_A^2)(q^2+k_B^2)}
\boldsymbol{\nabla}_{\mathbf{R}}[\boldsymbol{\mu}_{B}\cdot\mathbb{G}(iqR)\cdot\boldsymbol{\mu}_{A}].\label{FBdist}
\end{split}
\end{equation}
\end{widetext}

Note that, as anticipated after Eq.(\ref{Force}), the conservative vdW forces cannot be written in the form $-\boldsymbol{\nabla}_{\mathbf{R}}
\langle W_{A,B}\rangle_{T}/2$ due to the functional dissymmetry of the time-dependent terms proportional to 
$\textrm{Re}\mathbb{G}(k_{A,B} R)\textrm{Im}\mathbb{G}(k_{A,B} R)$ --for comparison, see  also the expressions for $\langle W_{A,B}\rangle_{T}$ in the Appendix.
Comparing equations (\ref{FAdist}) and (\ref{FBdist}), we observe that the only term which is common to both expressions is that involving off-resonant photons, 
which is proportional to $(2e^{-\Gamma_{A} T}-1)$. 
That implies that it changes sign at an observation time 
$T\approx\log{2}/\Gamma_{A}$. As for the rest of the terms, some fully resonant 
and semi-resonant terms, either stationary or oscillating at frequency $\Delta_{AB}$, differ in sign. Those terms constitute non-reciprocal forces and 
amount to a net force on the two-atom system. The stationary non-reciprocal forces were shown in Ref.\cite{My_Net_PRA} to  result from the excess of momenta stored in the virtual photons which mediate the resonant 
interaction in the processes depicted by 
diagrams (a) and (g). In addition, the slowly oscillating non-reciprocal forces arise after a sudden excitation only, and their associated 
momentum variation is supplied by the resonant photons of diagrams (a,c,d,e,f). Assuming that $|\Delta_{AB}|\ll\omega_{A,B}$ for the oscillating forces to be observable, 
the net force on the atomic system reads  
\begin{widetext}
\begin{align}\label{forcedist}
\langle&\mathbf{F}_{A}+\mathbf{F}_{B}\rangle_{T}\simeq\frac{8e^{-\Gamma_A T}k_{A}^{4}}{\epsilon_0^2\hbar}\frac{\omega_{B}}{\omega_{A}^{2}-\omega_{B}^{2}}
\boldsymbol{\mu}_{A}\cdot\textrm{Im}\mathbb{G}(k_A R)\cdot\boldsymbol{\mu}_{B}\boldsymbol{\nabla}_{\mathbf{R}}\left[\boldsymbol{\mu}_{B}\cdot
\textrm{Im}\mathbb{G}(k_A R)\cdot\boldsymbol{\mu}_{A}\right]\nonumber\\
&+\frac{2 e^{-(\Gamma_A+\Gamma_{B})T/2}k_{B}^2}{\epsilon_0^2\hbar\Delta_{AB}}\Bigr\{\boldsymbol{\mu}_{A}\cdot\textrm{Re}\mathbb{G}(k_B R)\cdot\boldsymbol{\mu}_{B}
\boldsymbol{\nabla}_{\mathbf{R}}\Bigl[
k_{B}^{2}\boldsymbol{\mu}_{B}\cdot\textrm{Re}\mathbb{G}(k_B R)\cdot\boldsymbol{\mu}_{A}
-k_{A}^{2}\boldsymbol{\mu}_{A}\cdot\textrm{Re}\mathbb{G}(k_A R)\cdot\boldsymbol{\mu}_{B}\Bigr]\nonumber\\
&-\boldsymbol{\mu}_{A}\cdot\textrm{Im}\mathbb{G}(k_B R)\cdot\boldsymbol{\mu}_{B}
\boldsymbol{\nabla}_{\mathbf{R}}\Bigl[k_{B}^{2}\boldsymbol{\mu}_{B}\cdot\textrm{Im}\mathbb{G}(k_B R)\cdot\boldsymbol{\mu}_{A}
+k_{A}^{2}\boldsymbol{\mu}_{A}\cdot\textrm{Im}\mathbb{G}(k_A R)\cdot\boldsymbol{\mu}_{B}\Bigr]\Bigr\}\cos{(\Delta_{AB}T)}\nonumber\\
&-\frac{2 e^{-(\Gamma_A+\Gamma_{B})T/2}k_{B}^2}{\epsilon_0^2\hbar\Delta_{AB}}\Bigr\{\boldsymbol{\mu}_{A}\cdot\textrm{Re}\mathbb{G}(k_B R)\cdot\boldsymbol{\mu}_{B}
\boldsymbol{\nabla}_{\mathbf{R}}\Bigl[
k_{B}^{2}\boldsymbol{\mu}_{B}\cdot\textrm{Im}\mathbb{G}(k_B R)\cdot\boldsymbol{\mu}_{A}
+k_{A}^{2}\boldsymbol{\mu}_{A}\cdot\textrm{Im}\mathbb{G}(k_A R)\cdot\boldsymbol{\mu}_{B}\Bigr]\nonumber\\
&+\boldsymbol{\mu}_{A}\cdot\textrm{Im}\mathbb{G}(k_B R)\cdot\boldsymbol{\mu}_{B}
\boldsymbol{\nabla}_{\mathbf{R}}\Bigl[k_{B}^{2}\boldsymbol{\mu}_{B}\cdot\textrm{Re}\mathbb{G}(k_B R)\cdot\boldsymbol{\mu}_{A}
-k_{A}^{2}\boldsymbol{\mu}_{A}\cdot\textrm{Re}\mathbb{G}(k_A R)\cdot\boldsymbol{\mu}_{B}\Bigr]\Bigr\}\sin{(\Delta_{AB}T)}\nonumber\\
&-\frac{2 e^{-(\Gamma_A+\Gamma_B) T/2} }{c^3 \epsilon_0^2 \hbar}\int_{0}^{\infty}\frac{dq}{\pi} 
\frac{(q^2-k_A k_B)q^2\boldsymbol{\mu}_{A}\cdot\mathbb{G}(iq R)\cdot\boldsymbol{\mu}_{B}}{(q^2+k_A^2)(q^2+k_B^2)}
\Bigl[\boldsymbol{\nabla}_{\mathbf{R}}[\omega^{2}_{B}\boldsymbol{\mu}_{A}\cdot\textrm{Re}
\mathbb{G}(k_B R)\cdot\boldsymbol{\mu}_B-\omega^{2}_{A}\boldsymbol{\mu}_{A}\cdot\textrm{Re}
\mathbb{G}(k_A R)\cdot\boldsymbol{\mu}_B]\nonumber\\
&\times\cos(\Delta_{AB}T)-\boldsymbol{\nabla}_{\mathbf{R}}[\omega_{B}^{2}\boldsymbol{\mu}_{A}\cdot\textrm{Im}\mathbb{G}(k_B R)
\cdot\boldsymbol{\mu}_B+\omega_{A}^{2}\boldsymbol{\mu}_{A}\cdot\textrm{Im}\mathbb{G}(k_A R)
\cdot\boldsymbol{\mu}_B]\sin(\Delta_{AB}T)\Bigr],
\end{align}
\end{widetext}

\noindent where the first non-oscillating term coincides with the net force for the case of an adiabatic excitation \cite{MePRA}. 

\begin{figure}[H]
	\centering
	\includegraphics[width=7cm,clip]{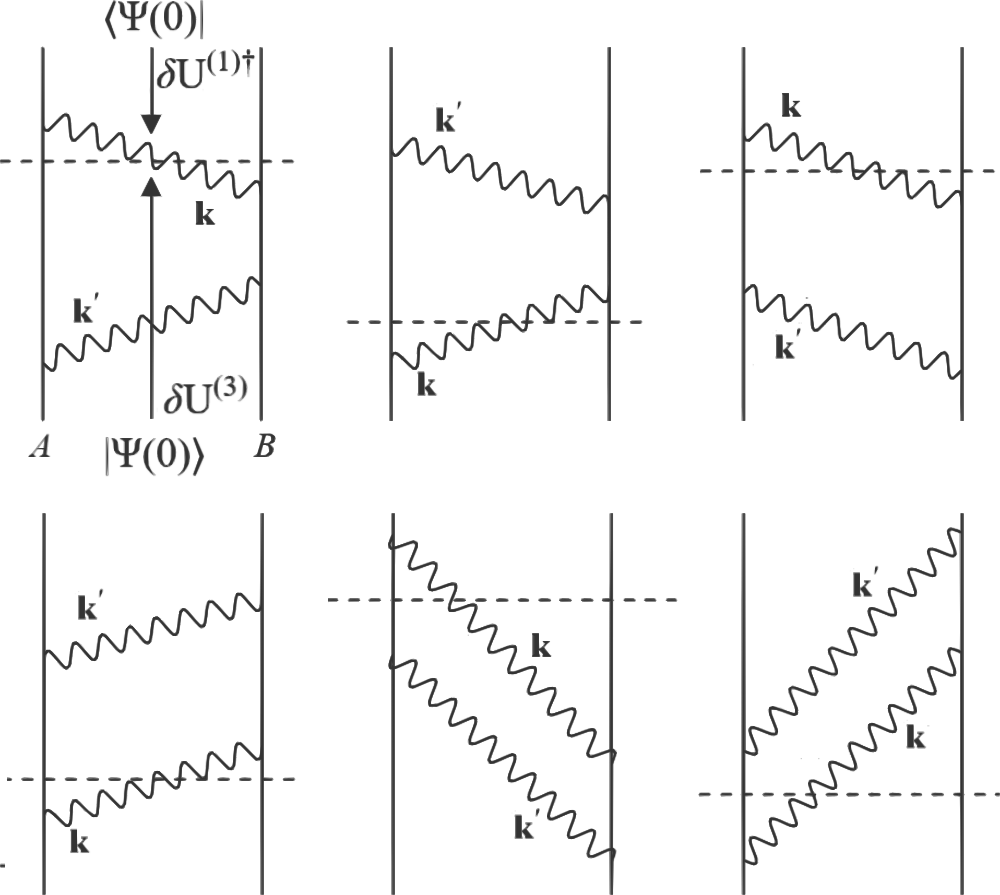}
	\caption{Diagrammatic representation of the processes which contribute to the one-photon directional emission rate, $d\Gamma_{\textrm{dir}}/d\Theta$.}\label{figure3}
\end{figure}

In what follows we study the directionality of one-photon spontaneous emission and show its relationship with the net  
force. Directionality is provided by the asymmetry in the emission rate of one of the resonant exchanged photons of the diagrams 
(a,c,d,e,f) depicted in Fig.\ref{figure3}.  Hence, the resultant formula for the directional emission rate as a function of the solid angle, 
$d\Gamma_{\textrm{dir}}/d\Theta$, 
is not invariant under parity inversion. Hence, the asymmetry is maximum along the interatomic axis. The evaluation of the one-photon emission diagrams in 
Fig.\ref{figure3} yields,
\begin{widetext}
	
	\begin{align}\label{Gammadiss}
		\frac{d\Gamma_{\textrm{dir}}}{d\Theta}= \frac{\boldsymbol{\mu}_A \cdot(\mathbb{I}-\hat{\mathbf{k}} \otimes \hat{\mathbf{k}}) 
		\cdot \boldsymbol{\mu}_B}{2( \pi \epsilon_0 \hbar)^2}
		\left\{\begin{array}{lcc} \frac{e^{-(\Gamma_A+\Gamma_B)T/2}k_B^5}{\Delta_{AB}} \Bigl[\cos(\Delta_{AB}T) 
		\left[\cos(k_B R \cos\theta) \boldsymbol{\mu}_A \cdot  \operatorname{Re}\mathbb{G}(k_B R) \cdot \boldsymbol{\mu}_B- \sin(k_B R \cos\theta) \right.\\ \left.
		 \times\boldsymbol{\mu}_A \cdot  \operatorname{Im}\mathbb{G}(k_B R) \cdot \boldsymbol{\mu}_B\right] -\sin(\Delta_{AB}T) 
		 \left[\cos(k_B R \cos\theta) \boldsymbol{\mu}_A \cdot  \operatorname{Im}\mathbb{G}(k_B R) \cdot \boldsymbol{\mu}_B 
		 \right. \\ \left.
		 +\sin(k_B R \cos\theta)\boldsymbol{\mu}_A \cdot  \operatorname{Re}\mathbb{G}(k_B R) \cdot \boldsymbol{\mu}_B\right]\Bigr]  + 
		 \frac{e^{-\Gamma_A T}k_A^5}{\Delta_{AB}}\sin(k_A R \cos\theta) \boldsymbol{\mu}_A \cdot  \operatorname{Im}\mathbb{G}(k_A R) \cdot 
		 \boldsymbol{\mu}_B \\ 
		 -e^{-(\Gamma_A+\Gamma_B)T/2} k_B^3\left[\cos(\Delta_{AB}T) \cos(k_B R \cos \theta)- \sin(\Delta_{AB}T) \sin(k_B R \cos \theta)\right]
			\\
		 \times \int_{0}^{\infty} c\frac{dq}{\pi} \frac{q^2(q^2-k_A k_B)}{(q^2+k_A^2)(q^2+k_B^2)}
		 \boldsymbol{\mu}_A \cdot \mathbb{G}(iqR) \cdot \boldsymbol{\mu}_B\quad\text{for } \cos \theta \in (0,1], \\ \,\ \\
		 \frac{e^{-(\Gamma_A+\Gamma_B)T/2}k_B^2k_A^3}{\Delta_{AB}} \Bigl[\cos(\Delta_{AB}T) \left[ \cos(k_A R \cos\theta) 
		 \boldsymbol{\mu}_A \cdot  \operatorname{Re}\mathbb{G}(k_B R) \cdot \boldsymbol{\mu}_B-\sin(k_A R \cos\theta)\right.\\
		\left.\times\boldsymbol{\mu}_A \cdot  \operatorname{Im}\mathbb{G}(k_B R) \cdot \boldsymbol{\mu}_B\right] 
		 -\sin(\Delta_{AB}T) \left[\cos(k_A R \cos\theta) \boldsymbol{\mu}_A \cdot  
		 \operatorname{Im}\mathbb{G}(k_B R) \cdot \boldsymbol{\mu}_B \right.\\
		 \left.+\sin(k_A R \cos\theta)\boldsymbol{\mu}_A \cdot  
		 \operatorname{Re}\mathbb{G}(k_B R) \cdot \boldsymbol{\mu}_B \right]\Bigr] +\frac{ e^{-\Gamma_A T}k_A^5}{\Delta_{AB}}
		 \sin(k_A R \cos\theta)\boldsymbol{\mu}_A \cdot  \operatorname{Im}\mathbb{G}(k_A R) \cdot \boldsymbol{\mu}_B
		\\
		 -e^{-(\Gamma_A+\Gamma_B)T/2} k_A^3\left[\cos(\Delta_{AB}T) \cos(k_A R \cos \theta)- \sin(\Delta_{AB}T) \sin(k_A R \cos \theta)\right]
			\\ 
		 \times \int_{0}^{\infty} c \frac{dq}{\pi} \frac{q^2(q^2-k_A k_B)}{(q^2+k_A^2)(q^2+k_B^2)}
		 \boldsymbol{\mu}_A \cdot \mathbb{G}(iqR) \cdot \boldsymbol{\mu}_B\quad\text{for } \cos \theta \in [-1,0).
		\end{array}
		\right.
	\end{align}
\end{widetext}
In this equation parity symmetry is manifestly broken by the difference between the terms defined in each interval of $\cos\theta$. In particular, those 
applicable to $\cos\theta\in(0,1]$ contribute to $\mathbf{F}_{A}$ upon integration of Eq.(\ref{Pdot}), while those for $\cos\theta\in[-1,0)$ 
contribute to $\mathbf{F}_{B}$, respectively.
 Under the condition $|\Delta_{AB}|\ll\omega_{A,B}$ and considering $\omega_{A}\simeq\omega_{B}$ for simplicity, we can write the time derivative of the transverse EM momentum as,
\begin{equation}\label{Pdot}
\langle\dot{\mathbf{P}}_{\perp}^{\gamma}\rangle_{T}\simeq\hbar k_{A}\int_{0}^{4\pi}d\Theta\hat{\mathbf{k}}\frac{d\Gamma_{\textrm{dir}}}{d\Theta}.
\end{equation}

\noindent Straight integration of this equation leads to Eq.(\ref{forcedist}) --up to two-photon emission terms-- but for a negative sign in front, proving that 
$\langle \mathbf{F}_{A}+\mathbf{F}_{B}\rangle_{T}=-\langle\dot{\mathbf{P}}_{\perp}^{\gamma}\rangle_{T}$ 
in agreement with the conservation of the momentum $\mathbf{K}$ defined in Eq.(\ref{K}).
\section{The identical atoms limit in the weak-interaction regime}\label{sec3}

We proceed to take the identical atoms limit upon the equations obtained in the previous section. That is, we consider $\omega_{B}\rightarrow\omega_{A}=\omega_{0}$,
$\Gamma_{B}\rightarrow\Gamma_{A}=\Gamma_{0}$, $\mathbf{\mu}_{A}=\mathbf{\mu}_{B}$. Note that, in order for the perturbative computations of Sec.\ref{sec2} to 
remain valid in this limit, the observation time $T$ must be small in comparison to the time that it takes for the excitation to be transferred from atom $A$ to 
atom $B$, i.e., $k_{0}^{2}\boldsymbol{\mu}_{A}\cdot$Re$\mathbb{G}(k_{0}R)\cdot\boldsymbol{\mu}_{B}\lesssim\hbar\epsilon_{0}/T$ 
\cite{Dicke,Craigbook}.  This is the weak-interaction 
regime, which implies that the original  atomic states are quasi-stationary despite the degeneracy of the system. 
In this limit, 
the vdW forces read
\begin{widetext}
\begin{align}
&\langle \mathbf{F}_A \rangle_{T}=-\frac{2\omega_{0}^{4}e^{-\Gamma_0 T}}{c^4 \epsilon_0^2 \hbar}T\:\Bigl[\boldsymbol{\mu}_{A}\cdot\textrm{Re}\mathbb{G}(k_0 R)\cdot\boldsymbol{\mu}_{B}
\boldsymbol{\nabla}_{\mathbf{R}}\left[\boldsymbol{\mu}_{B}\cdot\textrm{Im}\mathbb{G}(k_0 R)\cdot\boldsymbol{\mu}_{A}\right]+\boldsymbol{\mu}_{A}\cdot\textrm{Im}
\mathbb{G}(k_0 R)\cdot\boldsymbol{\mu}_{B}\boldsymbol{\nabla}_{\mathbf{R}}\left[\boldsymbol{\mu}_{B}\cdot\textrm{Re}\mathbb{G}(k_0 R)\cdot\boldsymbol{\mu}_{A}\right]\Bigr]\nonumber\\
&-\frac{2e^{-\Gamma_0 T}}{c^4 \epsilon_0^2 \hbar}\frac{\partial}{\partial\omega}\Bigl[\omega^{4}\boldsymbol{\mu}_{A}\cdot\textrm{Re}\mathbb{G}(k R)\cdot
\boldsymbol{\mu}_{B}\boldsymbol{\nabla}_{\mathbf{R}}\left[\boldsymbol{\mu}_{B}\cdot\textrm{Re}\mathbb{G}(k R)\cdot\boldsymbol{\mu}_{A}\right]-
\omega^{4}\boldsymbol{\mu}_{A}\cdot\textrm{Im}\mathbb{G}(k R)\cdot\boldsymbol{\mu}_{B}\boldsymbol{\nabla}_{\mathbf{R}}
\left[\boldsymbol{\mu}_{B}\cdot\textrm{Im}\mathbb{G}(k R)\cdot\boldsymbol{\mu}_{A}\right]\Bigr]_{\omega=\omega_{0}}\nonumber\\
&+\frac{\omega_{0}^{3}e^{-\Gamma_0 T}}{c^4 \epsilon_0^2\hbar}\Bigl[\boldsymbol{\mu}_{A}\cdot\textrm{Re}\mathbb{G}(k_0 R)\cdot\boldsymbol{\mu}_{B}
\boldsymbol{\nabla}_{\mathbf{R}}\left[\boldsymbol{\mu}_{B}\cdot\textrm{Re}\mathbb{G}(k_0 R)\cdot\boldsymbol{\mu}_{A}\right]-\boldsymbol{\mu}_{A}\cdot\textrm{Im}
\mathbb{G}(k_0 R)\cdot\boldsymbol{\mu}_{B}\boldsymbol{\nabla}_{\mathbf{R}}\left[\boldsymbol{\mu}_{B}\cdot\textrm{Im}\mathbb{G}(k_0 R)\cdot\boldsymbol{\mu}_{A}\right]\Bigr]\nonumber\\
&-\frac{2\omega_{0}^{2}e^{-\Gamma_0 T}}{c^3 \epsilon_0^2 \hbar}\boldsymbol{\nabla}_{\mathbf{R}}[\boldsymbol{\mu}_{A}\cdot\textrm{Re}\mathbb{G}(k_0 R)
\cdot\boldsymbol{\mu}_B]\int_{0}^{\infty}\frac{dq}{\pi}\frac{(q^2-k_0^2)q^2\boldsymbol{\mu}_{A}\cdot\mathbb{G}(iq R)\cdot\boldsymbol{\mu}_{B}}{(q^2+k_0^2)^{2}}\nonumber\\
&+\frac{4\omega_{0}^{2}\left(1-2e^{-\Gamma_{0} T}\right)}{c^3\epsilon_{0}^{2}\hbar}\int_{0}^{\infty}\frac{dq}{\pi}
\frac{q^4 \boldsymbol{\mu}_{A}\cdot\mathbb{G}(iqR)\cdot\boldsymbol{\mu}_{B} }{(q^2+k_0^2)^{2}}
\boldsymbol{\nabla}_{\mathbf{R}}[\boldsymbol{\mu}_{B}\cdot\mathbb{G}(iqR)\cdot\boldsymbol{\mu}_{A}],\label{FA0}
\end{align}
\begin{align}
&\langle \mathbf{F}_B \rangle_{T}=\frac{2\omega_{0}^{4}e^{-\Gamma_0 T}}{c^4 \epsilon_0^2 \hbar}T\:\Bigl[\boldsymbol{\mu}_{A}\cdot\textrm{Re}
\mathbb{G}(k_0 R)\cdot\boldsymbol{\mu}_{B}
\boldsymbol{\nabla}_{\mathbf{R}}\left[\boldsymbol{\mu}_{B}\cdot\textrm{Im}\mathbb{G}(k_0 R)\cdot\boldsymbol{\mu}_{A}\right]-\boldsymbol{\mu}_{A}\cdot\textrm{Im}
\mathbb{G}(k_0 R)\cdot\boldsymbol{\mu}_{B}\boldsymbol{\nabla}_{\mathbf{R}}\left[\boldsymbol{\mu}_{B}\cdot\textrm{Re}\mathbb{G}(k_0 R)\cdot\boldsymbol{\mu}_{A}\right]\Bigr]\nonumber\\
&+\frac{2e^{-\Gamma_0 T}\omega_{0}^{2}}{c^4 \epsilon_0^2 \hbar}\Bigl[\frac{\partial}{\partial \omega} \left[\omega^{2}\boldsymbol{\mu}_{A}\cdot\textrm{Re}\mathbb{G}(k R)\cdot
\boldsymbol{\mu}_{B} \right]_{\omega=\omega_{0}} \boldsymbol{\nabla}_{\mathbf{R}}\left[\boldsymbol{\mu}_{B}\cdot\textrm{Re}\mathbb{G}(k_{0} R)\cdot\boldsymbol{\mu}_{A}\right]
\nonumber\\&
+\frac{\partial}{\partial \omega} \left[\omega^{2}\boldsymbol{\mu}_{A}\cdot\textrm{Im}\mathbb{G}(k R)\cdot
\boldsymbol{\mu}_{B} \right]_{\omega=\omega_{0}}\boldsymbol{\nabla}_{\mathbf{R}}
\left[\boldsymbol{\mu}_{B}\cdot\textrm{Im}\mathbb{G}(k_{0} R)\cdot\boldsymbol{\mu}_{A}\right]\Bigr]\nonumber\\
&-\frac{\omega_{0}^{3}e^{-\Gamma_0 T}}{c^4 \epsilon_0^2\hbar}\Bigl[\boldsymbol{\mu}_{A}\cdot\textrm{Re}\mathbb{G}(k_0 R)\cdot\boldsymbol{\mu}_{B}
\boldsymbol{\nabla}_{\mathbf{R}}\left[\boldsymbol{\mu}_{B}\cdot\textrm{Re}\mathbb{G}(k_0 R)\cdot\boldsymbol{\mu}_{A}\right]+\boldsymbol{\mu}_{A}\cdot\textrm{Im}
\mathbb{G}(k_0 R)\cdot\boldsymbol{\mu}_{B}\boldsymbol{\nabla}_{\mathbf{R}}\left[\boldsymbol{\mu}_{B}\cdot\textrm{Im}\mathbb{G}(k_0 R)\cdot\boldsymbol{\mu}_{A}\right]\Bigr]\nonumber\\
&+\frac{2\omega_{0}^{2}e^{-\Gamma_0 T}}{c^3 \epsilon_0^2 \hbar}\boldsymbol{\nabla}_{\mathbf{R}}[\boldsymbol{\mu}_{A}\cdot\textrm{Re}\mathbb{G}(k_0 R)
\cdot\boldsymbol{\mu}_B]\int_{0}^{\infty}\frac{dq}{\pi}\frac{(q^2-k_0^2)q^2\boldsymbol{\mu}_{A}\cdot\mathbb{G}(iq R)\cdot\boldsymbol{\mu}_{B}}{(q^2+k_0^2)^{2}}\nonumber\\
&-\frac{4\omega_{0}^{2}\left(1-2e^{-\Gamma_{0} T}\right)}{c^3\epsilon_{0}^{2}\hbar}\int_{0}^{\infty}\frac{dq}{\pi}
\frac{q^4 \boldsymbol{\mu}_{A}\cdot\mathbb{G}(iqR)\cdot\boldsymbol{\mu}_{B} }{(q^2+k_0^2)^{2}}
\boldsymbol{\nabla}_{\mathbf{R}}[\boldsymbol{\mu}_{B}\cdot\mathbb{G}(iqR)\cdot\boldsymbol{\mu}_{A}],\label{FB0}
\end{align}
\end{widetext}

\begin{figure}[H]
\includegraphics[width=8.5cm,clip]{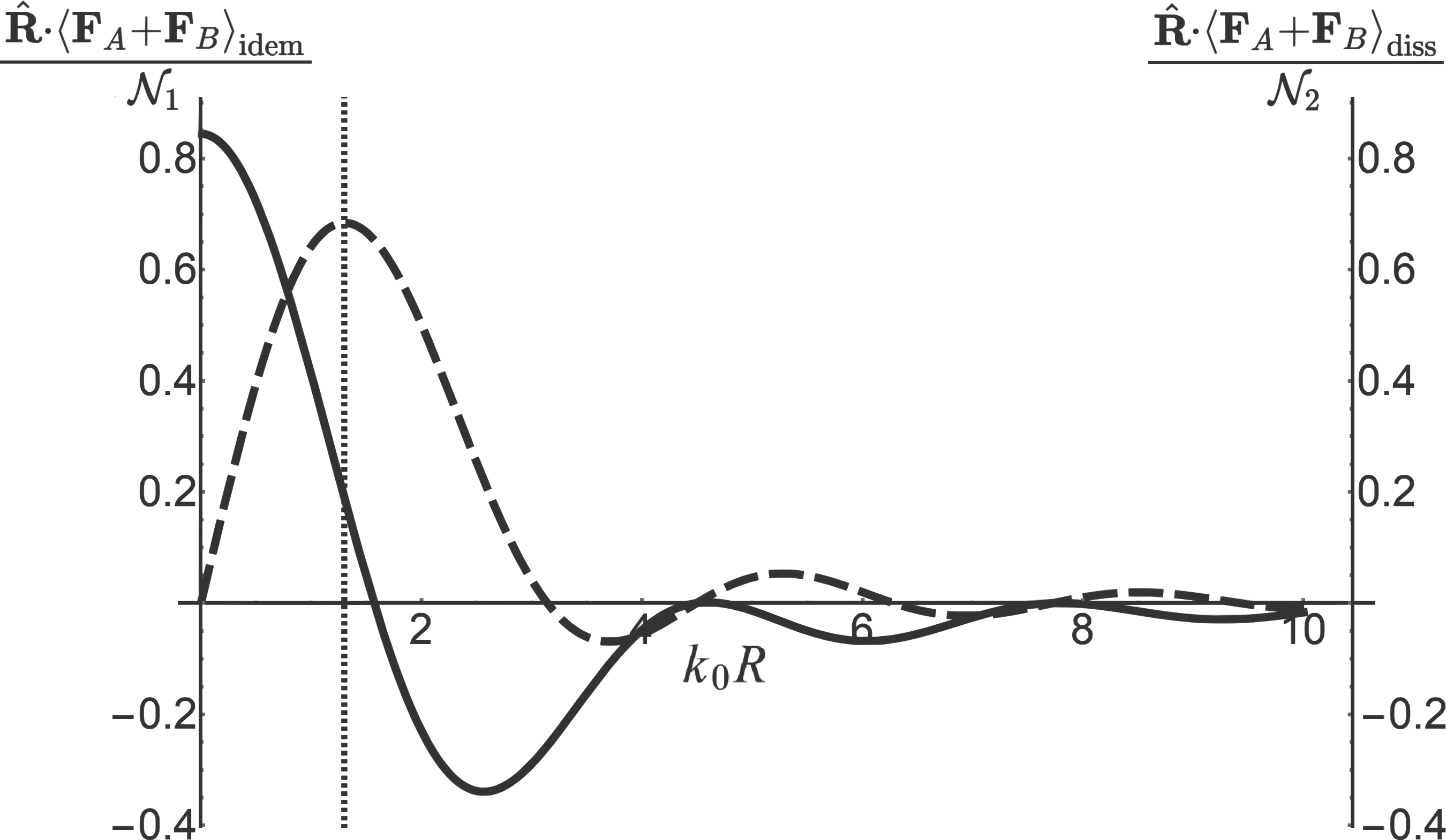}
\caption{Graphical representation of the net force on a binary system of identical atoms according to Eq.(\ref{forceidem}) as a function of $k_{0}R$, 
$\langle\mathbf{F}_{A}+\mathbf{F}_{B}\rangle_{\textrm{idem}}$ --solid curve,
normalized to $\mathcal{N}_1=\frac{|\boldsymbol{\mu}_A|^2 |\boldsymbol{\mu}_B|^2 \omega_0^7 T}{10^{2}c^{7}\hbar\epsilon_{0}^{2}}$; and net force on a binary 
system of dissimilar atoms according to Ref.\cite{My_Net_PRA}, $\langle\mathbf{F}_{A}+\mathbf{F}_{B}\rangle_{\textrm{diss}}$ --dashed curve, normalized to 
$\mathcal{N}_2=\frac{|\boldsymbol{\mu}_A|^2 |\boldsymbol{\mu}_B|^2 \omega_A^7}{10^{2}c^{7}\hbar\epsilon_{0}^{2}\Delta_{AB}}$, with $\omega_{A}\approx\omega_{0}$.}\label{figureNet}
\end{figure}
\noindent and the net force upon the atomic system is
\begin{align}
&\langle \mathbf{F}_A +\mathbf{F}_B\rangle_{T}=-\frac{4e^{-\Gamma_0 T}}{c^4 \epsilon_0^2 \hbar}\label{forceidem}\\
&\times\Bigl\{\omega_{0}^{4}T\:\Bigl[\boldsymbol{\mu}_{A}\cdot\textrm{Re}\mathbb{G}(k_0 R)\cdot\boldsymbol{\mu}_{B}
\boldsymbol{\nabla}_{\mathbf{R}}\left[\boldsymbol{\mu}_{B}\cdot\textrm{Im}\mathbb{G}(k_0 R)\cdot\boldsymbol{\mu}_{A}\right]\nonumber\\
&-\omega_{0}^{4}\frac{\partial}{\partial\omega}\Bigl[
\boldsymbol{\mu}_{A}\cdot\textrm{Im}\mathbb{G}(k R)\cdot\boldsymbol{\mu}_{B}\boldsymbol{\nabla}_{\mathbf{R}}
\left[\boldsymbol{\mu}_{B}\cdot\textrm{Im}\mathbb{G}(k R)\cdot\boldsymbol{\mu}_{A}\right]\Bigr]_{\omega=\omega_{0}}\nonumber\\
&-\frac{5\omega_{0}^{3}}{2}\boldsymbol{\mu}_{A}\cdot\textrm{Im}
\mathbb{G}(k_0 R)\cdot\boldsymbol{\mu}_{B}\boldsymbol{\nabla}_{\mathbf{R}}\left[\boldsymbol{\mu}_{B}\cdot\textrm{Im}\mathbb{G}(k_0 R)\cdot\boldsymbol{\mu}_{A}\right]\nonumber\\
&+\omega_{0}^{3}\boldsymbol{\mu}_{A}\cdot\textrm{Re}
\mathbb{G}(k_0 R)\cdot\boldsymbol{\mu}_{B}\boldsymbol{\nabla}_{\mathbf{R}}\left[\boldsymbol{\mu}_{B}\cdot\textrm{Re}\mathbb{G}(k_0 R)\cdot\boldsymbol{\mu}_{A}\right]\Bigr\},\nonumber
\end{align}
which contains fully resonant terms only.

As for the one-photon directional emission rate, taking the identical atoms limit on Eq.(\ref{Gammadiss}), we arrive at
\begin{widetext}
\begin{align}\label{Gammaidem}
\frac{d\Gamma_{\textrm{dir}}}{d\Theta}&=\frac{-\boldsymbol{\mu}_A \cdot(\mathbb{I}-\hat{\mathbf{k}} \otimes \hat{\mathbf{k}}) \cdot \boldsymbol{\mu}_B}{2(\pi \epsilon_0 \hbar)^2}e^{-\Gamma_0 T} 
\Bigl\{T k_0^5\sin(k_0 R \cos \theta) \boldsymbol{\mu}_A \cdot \operatorname{Re} \mathbb{G}(k_0 R) \cdot \boldsymbol{\mu}_B\nonumber\\
&-k_{0}^{5}\sin(k_0 R \cos \theta)\frac{\partial}{\partial \omega}[\boldsymbol{\mu}_A \cdot \operatorname{Im} \mathbb{G}(k R) \cdot \boldsymbol{\mu}_B]_{\omega=\omega_0}
-2c^{-1}k_{0}^{4}\sin(k_0 R \cos \theta)\boldsymbol{\mu}_A \cdot \operatorname{Im} \mathbb{G}(k_{0}R) \cdot \boldsymbol{\mu}_B\nonumber\\
&+H(\cos\theta)\:\Bigl[3c^{-1}k_{0}^{4}[\cos(k_0 R \cos \theta)\boldsymbol{\mu}_A \cdot \operatorname{Re} \mathbb{G}(k_{0}R) \cdot \boldsymbol{\mu}_B-\sin(k_0 R \cos \theta)\boldsymbol{\mu}_A \cdot \operatorname{Im} \mathbb{G}(k_{0}R) \cdot \boldsymbol{\mu}_B]\nonumber\\
&-c^{-1}Rk_{0}^{5}\cos\theta[\sin(k_0 R \cos \theta)\boldsymbol{\mu}_A \cdot \operatorname{Re} \mathbb{G}(k_{0}R) \cdot \boldsymbol{\mu}_B+\cos(k_0 R \cos \theta)\boldsymbol{\mu}_A \cdot \operatorname{Im} \mathbb{G}(k_{0}R) \cdot \boldsymbol{\mu}_B]\Bigr]\Bigr\},
\end{align}
\end{widetext}
where $H$ is the Heaviside function. The terms of Eq.(\ref{Gammaidem}) are in 
correspondence with those in Eq.(\ref{forceidem}) --but for two-photon emission terms, such that
\begin{equation}\label{Pdoto}
\langle \mathbf{F}_A +\mathbf{F}_B\rangle_{T}=-\langle\dot{\mathbf{P}}_{\perp}^{\gamma}\rangle_{T}\simeq-\hbar k_{0}\int_{0}^{4\pi}d\Theta\hat{\mathbf{k}}\frac{d\Gamma_{\textrm{dir}}}{d\Theta}.
\end{equation}
Two-photon emission terms together with those terms 
proportional to $\omega_{0}^{3}$ in Eq.(\ref{forceidem}) are indeed negligible in comparison to the term linear in $T$. In Fig.\ref{figureNet} we represent 
the net force on a binary system of identical atoms as a function of the interatomic distance, once normalized as indicated. 
For simplicity, the dipole moments are chosen isotropic, $\mu_{A,B}^{x}=\mu_{A,B}^{y}=\mu_{A,B}^{z}$. Note that, in order to preserve the perturbative nature of our calculation, 
the following inequality must be satisfied, $24\pi\textrm{Tr}\{\textrm{Re}\:\mathbb{G}(k_{0}R)\}\lesssim k_{0}/\Gamma_{0}T$. Considering the lower bound value of this 
inequality,  $\Gamma_{0}T\sim1$, it implies for isotropic dipoles $k_{0}R\gtrsim1.3$, as 
indicated with the vertical straight line in Fig.\ref{figureNet}. That implies that the actual maximum value of the net force on identical atoms is achieved at 
$k_{0}R\approx2.5$. For comparison, we represent  the net force on a binary system of dissimilar atoms \cite{My_Net_PRA}, which is of the order of 
$1/\Delta_{AB}T$ times smaller. The force on dissimilar atoms presents a maximum at $k_{0}R\approx1.3$, which coincides approximately with the value at which the 
force on the identical atoms vanishes for the first time.

\section{Conclusions}\label{sec4}
In the first place, starting with the pertuvative time-dependent computation of the dipole-dipole interaction between two dissimilar atoms, up to two-photon 
exchange processes, with one of the atoms suddenly excited, we have shown that the dipole-dipole forces contain two components. Namely, conservative forces 
identifiable with the ordinary van-der-Waals forces; and non-conservative forces which derive from the time-variation of the longitudinal EM momentum. 
In contrast to previous quasi-stationary  computations we find that, generally, the time-dependent vdW forces cannot be written as the gradients of the 
expectation values of the  interaction potentials, but as the expectation values of the gradients of the interaction potentials only. As for the non-conservative 
forces, they will be computed in a separate publication \cite{PRAJulio2}.

Second, we have taken the identical atoms limit upon the perturbative expresions for the vdW forces on dissimilar atoms. That compels us to constraint ourselves to 
the weak-interaction regime. We find that, at leading order, the van-der-Waals forces are fully-resonant and grow linearly in time, being different on each atom. 
Besides, in addition to the familiar off-resonant vdW forces, which change direction at $T=\log{2}/\Gamma_{0}$, semi-resonant 
reciprocal forces arise --Eqs.(\ref{FA0}) and (\ref{FB0}). 
 The resultant net force on the two-atom system 
 is related to the directionality of spontaneous emission, which results from the violation of parity symmetry and is in agreement with 
 total momentum conservation --Eqs.(\ref{Gammaidem}) and (\ref{Pdoto}). 

Beyond the weak-interaction regime the calculation of the vdW forces between identical atoms becomes non-perturbative as a result of degeneracy. That implies that 
non-perturbative time-evolution propagators are to be computed \cite{Varfolomeev1,Varfolomeev2}. Their calculation will be addressed in a separate publication, 
together with a proposal for the experimental observation of the net force on a binary system of Rydberg atoms.

\acknowledgments
Financial support from grants MTM2014-57129-C2-1-P (MINECO) and VA137G18, BU229P18 (JCyL) is acknowledged.
\appendix

\section{Interaction energies}
In this Appendix we compile the expressions for the interaction energies on each atom. Their diagrammatic representations are analogous to those in 
Figs.\ref{figure1A}-\ref{figure2B}, but for the replacement of the operators $-\boldsymbol{\nabla}_{A}W_{A}$ and $-\boldsymbol{\nabla}_{B}W_{B}$ at the observation 
time $T$ with $W_{A}$ and $W_{B}$, respectively,
\begin{widetext}
\begin{align}
\langle W_A\rangle_{T}&= \frac{2\omega_A^4 e^{-\Gamma_A T}}{c^{4}\epsilon_{0}^{2}\hbar\Delta_{AB}} \Bigl[[\boldsymbol{\mu}_{A}\cdot\text{Re}\mathbb{G}(k_{A}R)
\cdot\boldsymbol{\mu}_{B}]^{2}-[\boldsymbol{\mu}_{A}\cdot\text{Im}\mathbb{G}(k_{A}R)\cdot\boldsymbol{\mu}_{B}]^{2}\Bigr]\nonumber\\
&-\frac{2\omega_B^4 e^{-(\Gamma_A+\Gamma_{B})T/2}}{c^{4}\epsilon_{0}^{2}\hbar\Delta_{AB}} \Bigl[[\boldsymbol{\mu}_{A}\cdot\text{Re}\mathbb{G}(k_{B}R)
\cdot\boldsymbol{\mu}_{B}]^{2}-[\boldsymbol{\mu}_{A}\cdot\text{Im}\mathbb{G}(k_{B}R)\cdot\boldsymbol{\mu}_{B}]^{2}\Bigr] \cos(\Delta_{AB}T)\nonumber\\
&+\frac{4\omega_B^4 e^{-(\Gamma_A+\Gamma_{B})T/2}}{c^{4}\epsilon_{0}^{2}\hbar\Delta_{AB}}\Bigl[\boldsymbol{\mu}_{A}\cdot\text{Re}\mathbb{G}(k_{B}R)
\cdot\boldsymbol{\mu}_{B}\boldsymbol{\mu}_{B}\cdot\text{Im}\mathbb{G}(k_{B}R)\cdot\boldsymbol{\mu}_{A}\Bigr]\sin(\Delta_{AB}T)\nonumber\\
& -\frac{2\omega_A^4 e^{-\Gamma_A T}}{c^{4}\epsilon_{0}^{2}\hbar(\omega_{A}+\omega_{B})} \Bigl[[\boldsymbol{\mu}_{A}\cdot\text{Re}\mathbb{G}(k_{A}R)
\cdot\boldsymbol{\mu}_{B}]^{2}-[\boldsymbol{\mu}_{A}\cdot\text{Im}\mathbb{G}(k_{A}R)\cdot\boldsymbol{\mu}_{B}]^{2}\Bigr]\nonumber\\
&+\frac{2\omega_B^2 e^{-(\Gamma_A+\Gamma_{B})T/2}}{c^{3}\epsilon_{0}^{2}\hbar}\Bigl[\boldsymbol{\mu}_{A}\cdot\text{Re}\mathbb{G}(k_{B}R)
\cdot\boldsymbol{\mu}_{B}\cos(\Delta_{AB}T)-\boldsymbol{\mu}_{A}\cdot\text{Im}\mathbb{G}(k_{B}R)
\cdot\boldsymbol{\mu}_{B}\sin{(\Delta_{AB}T)}\Bigr]\nonumber\\
&\times\int_{0}^{\infty}\frac{dq}{\pi} \frac{(q^2-k_A k_B) q^2}{(q^2+k_A^2)(q^2+k_B^2)}\boldsymbol{\mu}_{B}\cdot\mathbb{G}(iqR)\cdot\boldsymbol{\mu}_{A}
-\frac{4\omega_A \omega_B\left(2e^{-\Gamma_{A} T}-1\right)}{c^3\epsilon_{0}^{2}\hbar}\int_{0}^{\infty}\frac{dq}{\pi}
\frac{q^4[\boldsymbol{\mu}_{A}\cdot\mathbb{G}(iqR)\cdot\boldsymbol{\mu}_{B}]^{2}}{(q^2+k_A^2)(q^2+k_B^2)},\label{WA_dist}
\end{align}
\begin{align}
\langle W_B\rangle_{T}&= \frac{2\omega_A^4 e^{-\Gamma_A T}}{c^{4}\epsilon_{0}^{2}\hbar\Delta_{AB}} \Bigl[[\boldsymbol{\mu}_{A}\cdot\text{Re}\mathbb{G}(k_{A}R)
\cdot\boldsymbol{\mu}_{B}]^{2}+[\boldsymbol{\mu}_{A}\cdot\text{Im}\mathbb{G}(k_{A}R)\cdot\boldsymbol{\mu}_{B}]^{2}\Bigr]\nonumber\\
&-\frac{2\omega_{A}^{2}\omega_B^2 e^{-(\Gamma_A+\Gamma_{B})T/2}}{c^{4}\epsilon_{0}^{2}\hbar\Delta_{AB}} \Bigl[\boldsymbol{\mu}_{A}\cdot\text{Re}\mathbb{G}(k_{A}R)
\cdot\boldsymbol{\mu}_{B} \boldsymbol{\mu}_{A}\cdot\text{Re}\mathbb{G}(k_{B}R)
\cdot\boldsymbol{\mu}_{B}+\boldsymbol{\mu}_{A}\cdot\text{Im}\mathbb{G}(k_{A}R)\cdot\boldsymbol{\mu}_{B}\boldsymbol{\mu}_{A}\cdot\text{Im}\mathbb{G}(k_{B}R)\cdot\boldsymbol{\mu}_{B}\Bigr] \nonumber\\ &\times \cos(\Delta_{AB}T)\nonumber\\
&-\frac{2\omega_{A}^{2}\omega_B^2 e^{-(\Gamma_A+\Gamma_{B})T/2}}{c^{4}\epsilon_{0}^{2}\hbar\Delta_{AB}}\Bigl[\boldsymbol{\mu}_{A}\cdot\text{Re}\mathbb{G}(k_{A}R)
\cdot\boldsymbol{\mu}_{B}\boldsymbol{\mu}_{B}\cdot\text{Im}\mathbb{G}(k_{B}R)\cdot\boldsymbol{\mu}_{A}-\boldsymbol{\mu}_{A}\cdot\text{Im}\mathbb{G}(k_{A}R)
\cdot\boldsymbol{\mu}_{B}\boldsymbol{\mu}_{B}\cdot\text{Re}\mathbb{G}(k_{B}R)\cdot\boldsymbol{\mu}_{A}\Bigr]\nonumber\\
&\times\sin(\Delta_{AB}T) \nonumber\\
&-\frac{2\omega_A^4 e^{-\Gamma_A T}}{c^{4}\epsilon_{0}^{2}\hbar(\omega_{A}+\omega_{B})} 
\Bigl[[\boldsymbol{\mu}_{A}\cdot\text{Re}\mathbb{G}(k_{A}R)
\cdot\boldsymbol{\mu}_{B}]^{2}+[\boldsymbol{\mu}_{A}\cdot\text{Im}\mathbb{G}(k_{A}R)\cdot\boldsymbol{\mu}_{B}]^{2}\Bigr]\nonumber \\
&+\frac{2\omega_A^2 e^{-(\Gamma_A+\Gamma_{B})T/2}}{c^{3}\epsilon_{0}^{2}\hbar}\Bigl[\boldsymbol{\mu}_{A}\cdot\text{Re}\mathbb{G}(k_{A}R)
\cdot\boldsymbol{\mu}_{B}\cos(\Delta_{AB}T)+\boldsymbol{\mu}_{A}\cdot\text{Im}\mathbb{G}(k_{A}R)
\cdot\boldsymbol{\mu}_{B}\sin{(\Delta_{AB}T)}\Bigr]\nonumber\\
&\times\int_{0}^{\infty}\frac{dq}{\pi} \frac{(q^2-k_A k_B) q^2}{(q^2+k_A^2)(q^2+k_B^2)}\boldsymbol{\mu}_{B}\cdot\mathbb{G}(iqR)\cdot\boldsymbol{\mu}_{A}
-\frac{4\omega_A \omega_B\left(2e^{-\Gamma_{A} T}-1\right)}{c^3\epsilon_{0}^{2}\hbar}\int_{0}^{\infty}\frac{dq}{\pi}
\frac{q^4[\boldsymbol{\mu}_{A}\cdot\mathbb{G}(iqR)\cdot\boldsymbol{\mu}_{B}]^{2}}{(q^2+k_A^2)(q^2+k_B^2)}.\label{WB_dist} 
\end{align}
\end{widetext}
Straight comparison with Eqs.(\ref{FAdist}) and (\ref{FBdist}) reveals that 
$-\boldsymbol{\nabla}_{A,B}\langle W_{A,B}\rangle_{T}/2\neq-\langle\boldsymbol{\nabla}_{A,B}W_{A,B}\rangle_{T}/2=\langle\mathbf{F}_{A,B}\rangle_{T}$, up to 
two-photon exchange processes.

\end{document}